\def\rn{\noindent\parshape 2 0truecm 8.5truecm 0.3truecm 8.2truecm}

\def\rn{}
\def\nn#1 #2{#2. #1}				
\def\nnn#1 #2 #3{#2. #3. #1}			
\def\nnnn#1 #2 #3 #4{#2. #3. #4 #1}		
\def\nnnnn#1 #2 #3 #4 #5{#2. #3. #4 #5. #1}	
\def\dualand{ and\hbox{ }}				
\def\multiand{ and,\hbox{ }}				
\def\rf#1;#2;#3;#4;#5 {{\frenchspacing\par\rn#1, #3 {\bf #4}, #5 (#2). \par}}
\def\rfbook#1;#2;#3;#4;#5 {{\frenchspacing\par\rn#1, {\it #3} (#5, #4, #2).\par}}
\def\rfprep#1;#2;#3 {{\par\frenchspacing\rn#1, #3 (#2).\par}}

\def\etal{{\frenchspacing\it et al.}}

\def\beq#1{\begin{equation}\label{#1}}
\def\eeq{\end{equation}}
\def\beqa#1{\begin{eqnarray}\label{#1}}
\def\eeqa{\end{eqnarray}}

\def\fig#1{Figure~\ref{#1}}
\def\Fig#1{Figure~\ref{#1}}

\def\spose#1{\hbox to 0pt{#1\hss}}
\def\simlt{\mathrel{\spose{\lower 3pt\hbox{$\mathchar"218$}}
     \raise 2.0pt\hbox{$\mathchar"13C$}}}
\def\simgt{\mathrel{\spose{\lower 3pt\hbox{$\mathchar"218$}}
     \raise 2.0pt\hbox{$\mathchar"13E$}}}
\def\simpropto{\mathrel{\spose{\lower 3pt\hbox{$\mathchar"218$}}
     \raise 2.0pt\hbox{$\propto$}}}

\def\ed{\end{document}}

\def\Ob{\Omega_{\rm b}}
\def\Oc{\Omega_{\rm cdm}}
\def\Ok{\Omega_{\rm k}}
\def\Ol{\Omega_\Lambda}
\def\Om{\Omega_{\rm m}}
\def\On{\Omega_\nu}
\def\Od{\Omega_{dm}}
\def\ob{\omega_{\rm b}}
\def\ocdm{\omega_{\rm cdm}}

\def\on{\omega_\nu}
\def\od{\omega_{dm}}
\def\Cl{C_\l}

\def\ns{n_s}
\def\nt{n_t}
\def\As{A_s}
\def\At{A_t}
\def\zion{z_{\rm ion}}

\def\l{\ell}

\def\Cl{C_\ell}

\documentstyle[prl,aps,epsf]{revtex}
\begin{document}
\twocolumn[\hsize\textwidth\columnwidth\hsize\csname@twocolumnfalse\endcsname


\preprint{IASSNS-AST 97/666}

\title{New CMB constraints on the cosmic matter budget:
trouble for nucleosynthesis?}

\author{Max Tegmark}

\address{Dept. of Physics, Univ. of Pennsylvania, 
Philadelphia, PA 19104; max@physics.upenn.edu}

\author{Matias Zaldarriaga}

\address{Institute for Advanced Study, Princeton, NJ 08540; matiasz@ias.edu}

\date{Submitted to {\it\frenchspacing Phys. Rev. Lett.} April 30, 2000;
revised July 10; accepted July 17; published September 11}

\maketitle

\begin{abstract}
We compute the joint constraints on ten cosmological parameters
from the latest CMB measurements.
The lack of a significant second acoustic peak in the latest Boomerang
and Maxima
data favors models with more baryons than Big Bang nucleosynthesis
predicts, almost independently of what prior information is included.
The simplest flat inflation models with purely scalar scale-invariant fluctuations
prefer a baryon density
$0.022<h^2\Ob<0.040$ and a total nonbaryonic (hot + cold) dark matter
density $0.14<h^2\Od<0.32$ at 95\% confidence,
and allow reionization no earlier than $z\sim 30$.
\end{abstract}

\pacs{98.62.Py, 98.65.Dx, 98.70.Vc, 98.80.Es}

] 



One of the main challenges in modern cosmology is to refine and test
the standard model of structure formation by precision measurements
of its free parameters. 
The cosmic matter budget involves at least the four 
parameters
$\Ob$, $\Oc$, $\On$ and $\Ol$, which give the
percentages of critical density corresponding to
baryons, cold dark matter, massive neutrinos and 
vacuum energy. 
A ``budget deficit'' $\Ok\equiv 1-\Ob-\Oc-\On-\Ol$ manifests itself
as spatial curvature.
The description of the initial seed fluctuations predicted by inflation 
requires at least four parameters, 
the amplitudes $\As$ \& $\At$ 
and slopes $\ns$ \& $\nt$ of scalar and tensor fluctuations,
respectively.
Finally, the optical depth parameter $\tau$ 
quantifies when the first stars or quasars reionized
the Universe and the Hubble parameter $h$ gives
its current expansion rate.

\begin{figure}[phbt]
\vskip-1.2cm
\centerline{{\vbox{\epsfxsize=9.5cm\epsfbox{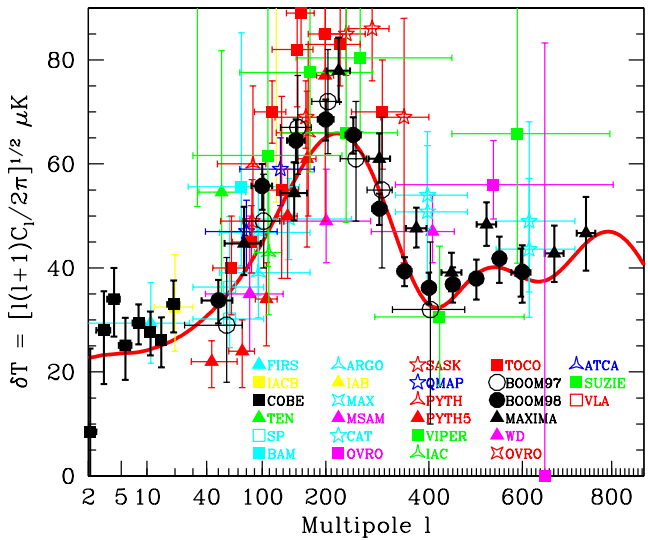}}}}
\vskip-0.9cm
\caption{
The 87 band power measurements used. 
The curve shows the simple inflationary model with
$\tau=\Ok=\On=r=0$, $\Ol=0.43$, $h^2\Oc=0.20$, $h^2\Ob=0.03$, $\ns=1$,
$h=0.63$.
Note that although we include the calibration uncertainties in our
analysis, they are not reflected by the plotted error bars.
}
\label{bestfitFig}
\end{figure}

During the past year or so, a number of papers
\cite{Lineweaver99,Dodel,Melchiorri,Lahav,10par,Efstathiou,LeDour,White} 
have used the measured  
cosmic microwave background (CMB) fluctuations
to constrain subsets of these parameters.
CMB data has improved dramatically since fluctuations
were first detected \cite{Smoot}.
The measurement of a first acoustic peak at the degree scale
\cite{Miller},
suggesting that the Universe is flat ($\Ok=0$), has now been
beautifully confirmed and improved by using the
ground-breaking high fidelity maps of the Boomerang
\cite{deBernardis} and Maxima \cite{Hanany} experiments.
As can be seen in \fig{bestfitFig},
perhaps the most important new information from Boomerang and
Maxima is
their accurate measurements of the angular power spectrum 
$\Cl$ on even smaller scales, out to multipole $\l\sim 600-800$.
The striking lack of a significant second acoustic peak places 
strong constraints on the cosmological parameters
\cite{White,deBernardis,Hu}, making a new full-fledged analysis 
of all the CMB data very timely.

In this {\it Letter}, we jointly constrain
the following 10 cosmological parameters: 
$\tau$, $\Ok$, $\Ol$, $n_s$, $n_t$, $A_s$, 
the tensor-to scalar ratio $r\equiv A_t/A_s$,  
and the physical matter densities 
$\ob\equiv h^2\Ob$, $\ocdm\equiv h^2\Oc$ and $\on\equiv h^2\On$.
The identity $h=\sqrt{(\ocdm+\ob+\on)/(1-\Ok-\Ol)}$
fixes the Hubble parameter.
We use the 10-dimensional grid method described in
\cite{10par}. In essence, this utilizes a technique for accelerating the
the CMBfast package \cite{cmbfast} by a factor around $10^3$ to 
compute theoretical power spectra on a grid in the 10-dimensional
parameter space, fitting these models to the data and then using cubic
interpolation of the resulting 10-dimensional likelihood function to 
marginalize it down to constraints on individual or pairs of
parameters. We use the 87 data points shown in
\fig{bestfitFig}, combining the 65 tabulated in \cite{10par}
with the 12 new Boomerang points \cite{deBernardis} and the 10 
new Maxima points \cite{Hanany}.

\begin{figure}[pbt]
\centerline{\hglue4.8cm\epsfxsize=13.7cm\epsffile{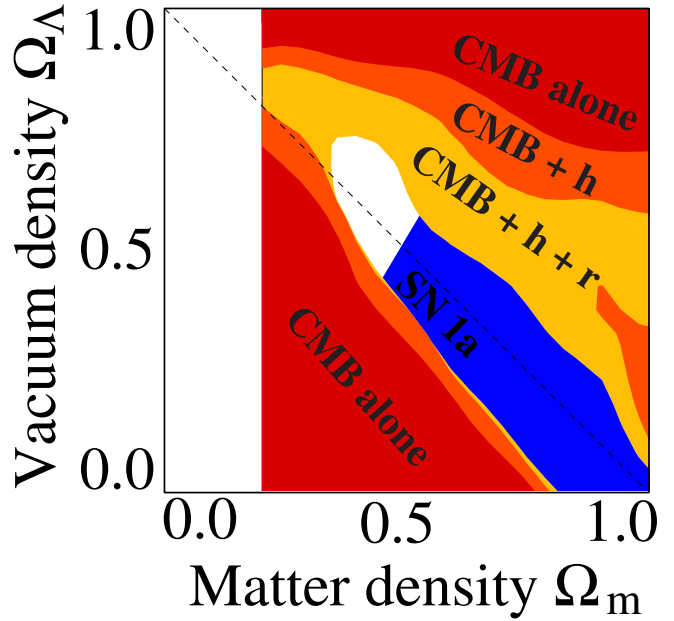}}
\vskip-4.5cm
\caption{
The regions in the $(\Om,\Ol)$-plane 
that are ruled out at 95\% are shown 
using 
(starting from the outside) 
no priors, 
the prior that $0.5<h<0.8$ (95\%), 
and the additional constraint $r=0$ 
The SN 1a constraints are from White [20] 
}
\label{OmOlFig}
\end{figure}
\begin{figure}[phbt]
\centerline{{\vbox{\epsfxsize=8.6cm\epsfbox{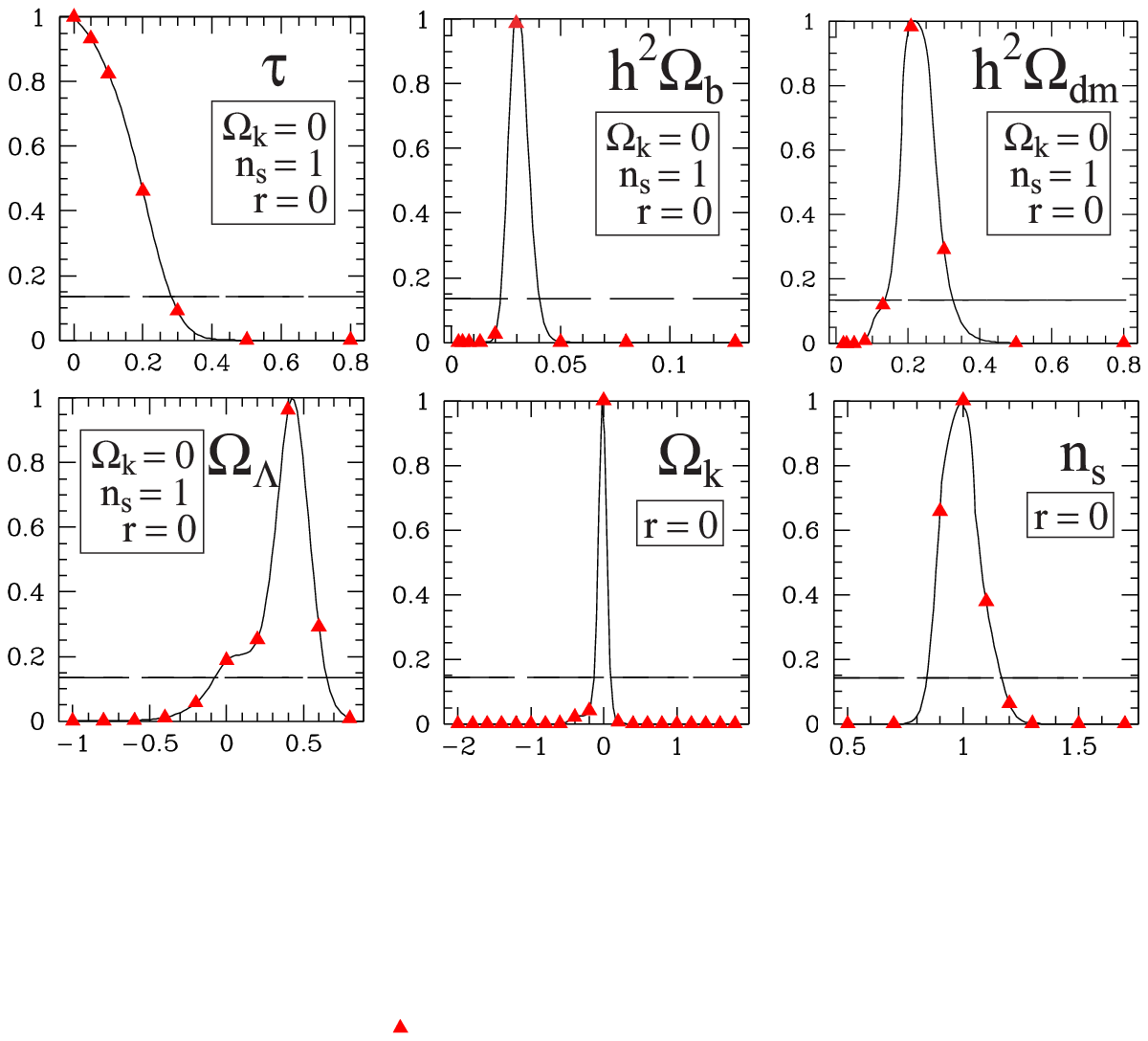}}}}
\vskip-2cm
\smallskip\smallskip
\caption{
Marginalized likelihoods assuming that 
$0.5<h<0.8$ (95\%) and the inflationary
priors specified in the boxes.
$2-\sigma$ limits are roughly 
where the curves drop beneath the dashed lines.
}
\label{1Dfig}
\end{figure}

Our 95\% confidence limits on the best constrained parameters
are summarized in Table 1.
\Fig{OmOlFig} shows that CMB alone suggests that the Universe
is either flat (near the diagonal line $\Om+\Ol=1$, where 
$\Om\equiv\Ob+\Oc+\On$)
or closed (upper right). These constraints come largely from the
location of the first peak, which is well-known to move to the right
if the curvature $\Ok$ is increased \cite{HuReview}.
Very closed models work only
because the first acoustic peak can also
be moved to the right by increasing the tilt $\ns$ 
or decreasing the matter density and bringing the
large-scale COBE signal back up with tensor fluctuations
(gravity waves)
\cite{9par,MelchiorriSazhin}. Galaxy clustering constraints disfavor
such strong blue-tilting, and Figures~\ref{OmOlFig}
and~\ref{1Dfig} show that closing this loophole by
barring gravity waves ($r=0$) favors curvature near zero
and $\ns$ near unity. This is a 
striking success for the oldest and simplest inflation models,
which make the three predictions $r\approx 0$, $\Ok\approx 0$ and $\ns\approx 1$ 
\cite{Steinhardt,Turner}.
Another important success for inflation is that the first 
peak is so narrow --- if the data had revealed the type of broad 
peak expected in many topological defect scenarios,
{\it none} of the models in our grid would have provided an acceptable fit.
Because of these tantalizing hints that ``back to basics''
inflation is correct, Table 1 and \fig{1Dfig} include results
assuming this inflation prior $r=\Ok=0, \ns=1$.

\begin{center}
\noindent
{\footnotesize
{\bf Table 1} -- Maximum-likelihood values and 95\% confidence limits.
The ``inflation prior'' for each parameter is indicated in boxes in \fig{1Dfig}.
$\Od\equiv\Oc+\On$.  A dash indicates that no 
limit was found, with the likelihood still above $e^{-2}$ at
the edge of our grid. Extrapolation would suggest a limit 
$\ns\simlt 1.75$.\\
}
\smallskip\smallskip

{
\begin{tabular}{|l|ccc|ccc|}
\hline
			&\multicolumn{3}{c|}{10 free parameters}
			&\multicolumn{3}{c|}{Inflation prior}\\
Quantity		&Min	&Best	&Max	&Min	&Best	&Max\\
\hline
$\tau$			&0.0	&0.0	&0.33	&0.0	&0.0	&0.28\\	
$h^2\Ob$		&.017	&.05	&$-$	&.022	&.03	&.040\\	
$h^2\Od$		&0.02	&$0.08$	&$-$	&0.14	&0.20	&0.32\\
$\Ol$			&$-$	&0.2	&0.80	&$-$0.16&.43	&0.65\\	
\hline		
$\Ok$			&$-$	&$-$0.6&0.13	&$-$0.13&0	&0.10\\	
$\ns$			&0.8	&1.5	&$-$	&0.84	&1.0	&1.17\\	
%
\hline
\end{tabular}
}
\end{center}

\bigskip

\begin{figure}[phbt]
\centerline{{\vbox{\epsfxsize=9.0cm\epsfbox{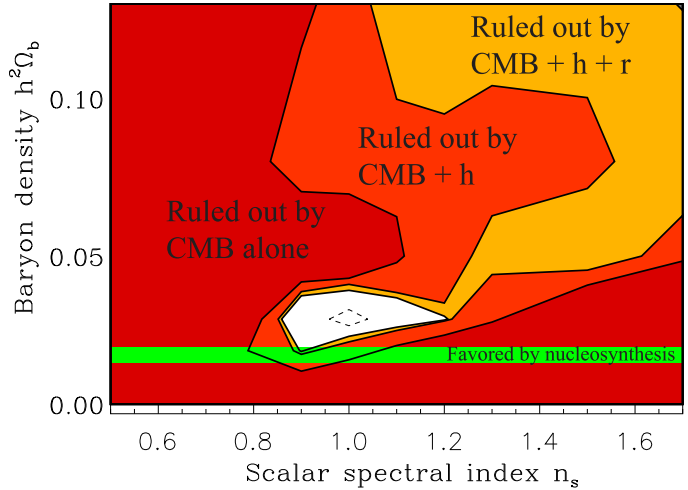}}}}
\smallskip\smallskip
\caption{
The regions in the $(\ns,\ob)$-plane 
that are ruled out at 95\% are shown 
using (starting from the outside) 
no priors, 
the prior $0.5 < h < 0.8$ (95\%),
the additional constraint $r=0$
and the additional constraints $\tau=0$ (dashed line).
The horizontal band shows the nucleosynthesis constraints
$\ob=0.019\pm 0.0024$.
}
\label{obnsFig}
\end{figure}

The constraints in Table 1 are seen to be much more interesting
than those before Boomerang and Maxima \cite{10par}, thanks to
new information on the scale of the second peak and beyond.
Cold dark matter and neutrinos have indistinguishable 
effects on the CMB except for very light neutrinos
(small $\on$), and the current data still lacks the precision
to detect this subtle difference.
The predicted height ratio of the first two peaks therefore depends
essentially on only three parameters \cite{White,Hu,observables}:
$\ns$, $\ob$ and $\od$, where the total dark matter density 
$\od\equiv\ocdm+\on$. Let us focus on the constraints on these parameters.
Increasing $\ob$ tends to boost the odd-numbered peaks (1, 3, etc.)
at the expense of even ones (2, 4, etc.) \cite{HuReview},
whereas increasing $\od$ suppresses all peaks
(see the CMB movies at $www.hep.upenn.edu/{\sim}max$ or
$www.ias.edu/{\sim}whu$).
The low second peak can therefore 
be fit by either 
decreasing the tilt $\ns$ or by increasing the baryon
density $\ob$ \cite{White,Hu} compared to the usually assumed values
$\ns\approx 1$, $\ob\approx 0.02$.
As illustrated in \fig{obnsFig}, this conclusion is essentially
independent of what priors are assumed.
However, reducing $\ns$ below 0.9 is seen to make things worse
again, as the first peak becomes too low relative 
to the COBE-normalization.

In short, there are two very simple ways of explaining the
lack of a prominent second acoustic peak: more baryons or a
red-tilted spectrum \cite{White,Hu}. However, as we will now 
discuss, both of these solutions have problems of their own.

\begin{figure}[phbt]
\centerline{{\vbox{\epsfxsize=9.0cm\epsfbox{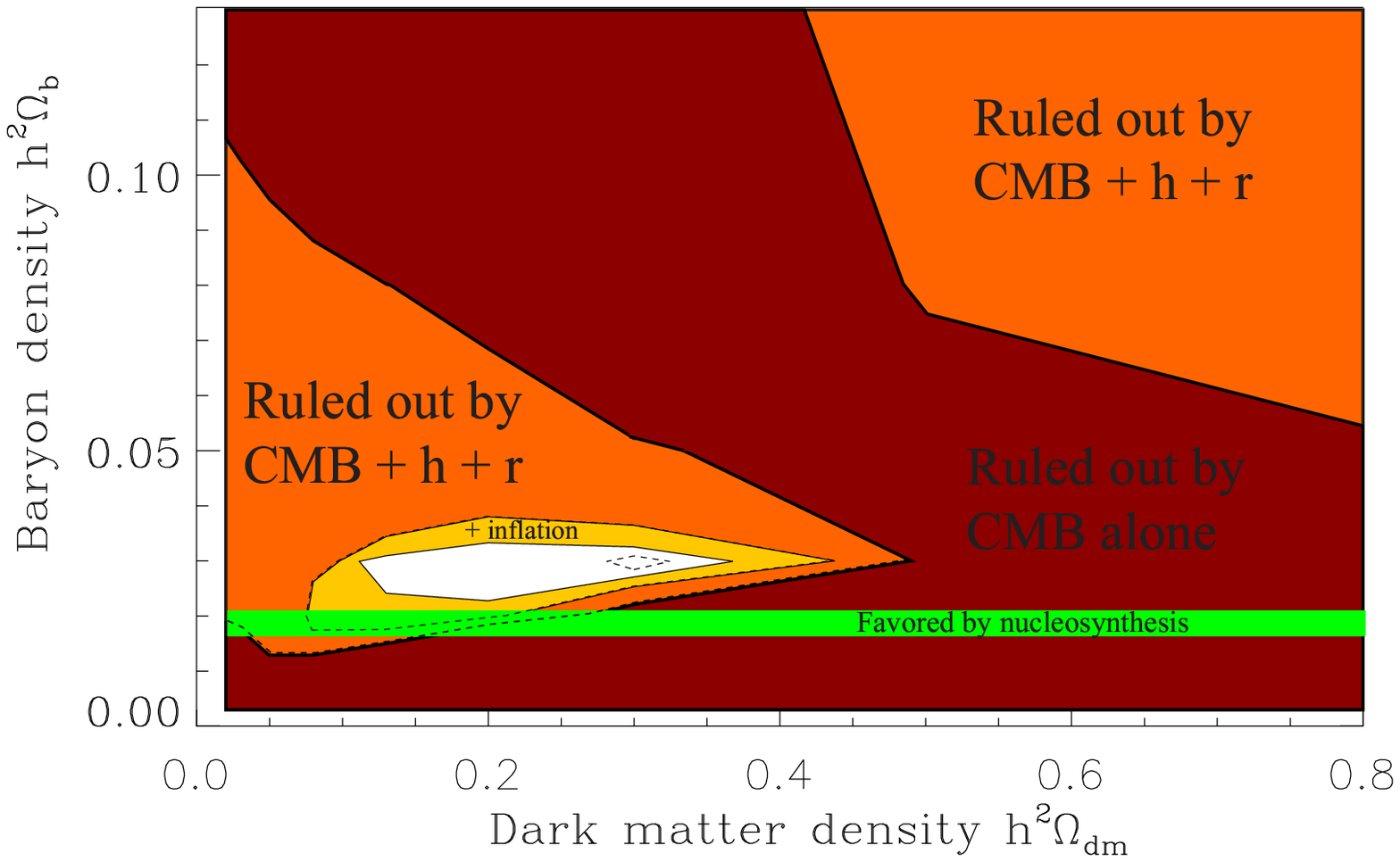}}}}
\smallskip\smallskip
\caption{
Similar to \fig{obnsFig}, but for for the $(\od,\ob)$-plane.
}
\label{ocobFig}
\end{figure}

\Fig{ocobFig} shows that when more baryons are added, 
more dark matter is needed to keep the first peak height constant.
When the tilt $\ns$ is fixed by the inflation prior, the constraints
on the remaining two parameters $\od$ and $\ob$ are seen to 
become quite tight. Intriguingly, the preferred baryon fraction is
of the same order as preferred by Big Bang nucleosynthesis,
but nonetheless higher than the tight nucleosynthesis error bars 
\cite{Tytler,Burles} $\ob=0.019\pm 0.0024$ allow. Even if the
nucleosynthesis error bars have somehow been underestimated so 
that $\ob\simgt 0.023$ as required
by the CMB data plus simple inflation is allowed, this solution may conflict
with other astrophysical constraints. For instance, X-ray observations of
clusters of galaxies can be used to determine the ratio of baryons to dark
matter \cite{Evrard,Mohr}, and $\ob=0.03$ 
can only be reconciled with these observations by
having $\Omega_{m}\simgt 0.7$ which would conflict with the supernova 1a
results and other estimates of the dark matter density \cite{Bahcall}.

On the other hand, the tilt solution is no panacea either.
In a class of popular inflationary models known as
power law inflation, 
the amplitude of the tensor component is approximately
related to the tilt of the scalar spectrum, 
$r\sim 7(1-n_s)$ \cite{Liddle}.
If we choose to fit the data by lowering the tilt to
$n_s=0.9$, this would 
raise the COBE-normalization by $70\%$.
Models that match the COBE normalization therefore make the first 
peak too low by a factor of 1.7 in power, which is ruled out by the data. 
In other words, imposing $r\sim 7(1-n_s)$ (which we have {\it not} done
in our analysis)
would exclude
$\ob$ as low as 0.02.
Thus the simple tilt solution does not work for all inflation models.

Could the apparent problem be a mere statistical fluke?
It would certainly be premature to claim a rock-solid discrepancy between 
CMB and nucleosynthesis plus power law inflation. 
The $\chi^2$-value for
the best fit inflation model with $\ob=0.02$ is still statistically 
acceptable
($\chi^2\approx 81$
for 87 degrees of freedom
reduced by about 5 effective parameters). 
However, serious discrepancies
in peak heights tend to get statistically diluted by the swarm of
points with large error bars at lower $\l$ that agree with 
most anything reasonable (indeed, $\chi^2$ drops down to 
71 for $\ob=0.03$ and as low as 68 without any priors),
and the relative likelihood rises sharply with 
$\ob$ regardless of what priors are imposed. 
To assess the sensitivity of the results to the choice of data, we therefore
repeated our entire analysis for the following cases:
(a) using all the data except Maxima and (b)
using only
COBE and the new Boomerang data. 
Omitting Maxima removed the ``CMB only'' and ``CMB+$h$'' exclusion regions
that are seen to protrude in from the left in \fig{obnsFig}.
This is because the Maxima points place an upper limit
on the height of (the left part of) the third peak, effectively
giving an upper limit on the baryon density.
Dropping Maxima also loosened the upper limit on $\od$ somewhat 
and marginally weakened the bounds on $\Ok$ and $\Ol$.
The other constraints were essentially unaffected.
Most importantly, the {\it lower} bound on 
$\ob$ seen in \fig{obnsFig} remained unchanged, since it comes
from the low ratio of the 2nd to 1st peak heights 
\cite{observables}.

Although our inclusion of the
10\% uncertainly in the Boomerang's calibration (20\% in power) was not very 
important in our full analysis, as the fitting
procedure de facto calibrated Boomerang off of other experiments,
this substantially degraded the results of our COBE + Boomerang analysis.
We therefore repeated it three more times, without the calibration
error but multiplying the Boomerang points 
by 0.9, 1.0 and 1.1, respectively. 
The results were quite similar to those using all the data, as expected from
the experimental concordance seen in \fig{bestfitFig}. 
However, most constraints got slightly tighter, consistent with the
above-mentioned $\chi^2$ dilution hypothesis.
Rather than go away, the baryon problem became exacerbated:
the 95\% inflationary lower limit on $\ob$ was tightened from $0.024$
to 0.027 with $\chi^2=12$ (with a total of 20 Boomerang + COBE
points and 4 free parameters). In contrast, the tilt
solution gave $\chi^2=22$, and higher still when the 
Boomerang normalization was raised or lowered by 10\%. 
Most strikingly, in the COBE+Boomerang version of \fig{obnsFig},
$\ob$ is not permitted to be low enough to agree with nucleosynthesis
for {\it any} value of the tilt $\ns$, so the tilt solution may have
worked using all the data merely because of the above-mentioned
dilution effect.

Can the baryon problem be explained by inaccuracies in our numerical
method? 
The correlations between the Boomerang points (which we could not include
since the have not yet been made public) are reportedly very small 
\cite{deBernardis}.
Although a range of approximations are involved as detailed
in \cite{10par}, for instance in the likelihood calculation, it appears
unlikely that such inaccuracies are large enough to have a major
impact on the lower bound on $\ob$. Perhaps the best indication of this
is that a number of independent analyses 
\cite{Lange,Bambi,observables,Lesgourgues}
have been made available 
since this paper was originally submitted, using a wide range
of computational techniques, and they all favor baryon fractions in
excess of the current nucleosynthesis prediction. 
More baryons also solve some older problems \cite{WhiteBaryons}.


A number of other ways out have been proposed \cite{White,Hu}, 
ranging from mechanisms for delaying standard recombination
to time-variation of physical constants
and severe mis-estimates of the Boomerang beam width.
However, these explanations are all of a highly speculative nature.
An excellent way to clear up this mystery will be to search 
for a third acoustic peak, which is boosted by more baryons
but suppressed by most of the other proposed remedies.

Apart from the matter budget, Table 1 and
\fig{1Dfig} also show that the CMB data provides 
perhaps the first meaningful upper
limit on $\tau$, the optical depth due to reionization 
(compare \cite{10par,LiddleTau}).
Since $\tau\propto h\Ob\zion^{3/2}\Om^{-1/2}$ 
if the redshift of reionization $\zion\gg 1$, 
our lower limit $\ob>0.024$ 
combined with our upper limit
$\tau<0.35$ give the constraint 
$\zion\simlt 49 h^{2/3}\Om^{1/3}$, 
or $\zion\simlt 28$ for $h=\Ol=0.7$.
This is compatible with the range $\zion=8-20$ 
favored by numerical simulations, 
but challenges more extreme models.


In conclusion, the new Boomerang results look like a 
triumph for the simplest possible inflationary model
but for one rather large fly in the ointment: the
lack of a significant second acoustic peak 
suggests that we may need to abandon either
a popular version of inflation,
the current nucleosynthesis constraints, or some even more
cherished assumption.
In answering one question, Boomerang has raised another.
Its answer is likely to lie in the 
third peak, and the race to reach it has now begun.

\bigskip
The authors wish to thank 
John Beacom,
Kevin Cahill,
Ang\'elica de Oliveira-Costa, Mark Devlin, 
Andrew Hamilton, David Hogg, Wayne Hu, Lam Hui,
William Kinney, Andrew Liddle, 
Dominik Schwarz,
Paul Steinhardt and
Ned Wright
for helpful comments and discussions.
Support for this work was provided by
NSF grant AST00-71213, 
NASA grant NAG5-9194 and 
Hubble Fellowship HF-01116.01-98A from 
STScI, operated by AURA, Inc. 
under NASA contract NAS5-26555.



\end{document}